# Generation of the squeezed state with an arbitrary complex amplitude distribution

Long Ma, Hui Guo, Hengxin Sun , Kui Liu, Bida Su, and Jiangrui Gao,

*State Key Laboratory of Quantum Optics and Quantum Optics Devices, Institute of Opto-Electronics,*
*Collaborative Innovation Center of Extreme Optics,*
*Shanxi University, Taiyuan, Shanxi 030006, China.*
*Science and Technology on Optical Radiation Laboratory, Beijing Institute of Environmental Features,*
*Beijing 100854, China*
*Corresponding author: liukui@sxu.edu.cn; jrgao@sxu.edu.cn*

**Abstract:** The squeezed state is important in quantum metrology and quantum information. The most effective generation tool known is the optical parametric oscillator (OPO). Currently, only the squeezed states of lower-order spatial modes can be generated by an OPO. However, the squeezed states of higher-order complex spatial modes are more useful for applications such as quantum metrology, quantum imaging and quantum information. A major challenge for future applications is efficient generation. Here, we use cascaded phase-only spatial light modulators to modulate the amplitude and phase of the incident fundamental mode squeezed state. This efficiently generates a series of squeezed higher-order Hermite–Gauss modes and a squeezed arbitrary complex amplitude distributed mode. The method may yield new applications in biophotonics, quantum metrology and quantum information processing.

## 1. Introduction

Continuous variable (CV) squeezed states attract much attention for their uses in quantum information processing, such as quantum communication[1], quantum computation[2], and quantum metrology[3,4]. At present, the most effective generation tool is an optical parametric oscillator (OPO). Traditionally, most OPOs operate in the fundamental mode. Recently, there has been growing interest in higher-order modes such as Hermite–Gauss (HG) and Laguerre–Gauss (LG) modes owing to their complex intensity and phase patterns, and different modes are used as independent variables. Their applications include rotation measurement with LG modes [5,6], micro-displacement measurement with HG modes [7,8], and thermal noise mitigation from fluctuations of mirror surfaces in coatings and substrates using the LG33 mode in gravitational wave interferometers [9, 10].

Traditionally, spatial mode squeezed states are generated by higher-order mode OPO. Quadrature squeezing of HG00, HG10, and HG20 modes and quadrature entanglement of first-order LG modes have been generated in a type-I OPO [11,12], and the higher-order mode squeezing or entanglement was enhanced by an optimized pump mode [13-18]. Moreover, the CV hyper-entanglement state, wherein both spin and orbital angular momenta are entangled, was realized in a multimode type-II OPO [19,20]. A specially designed OPO can also generate multimode squeezing and entanglement [21-26]. However, the studies above just produced low-order mode squeezing. There is no report on HG mode squeezed light generation higher than third order. As the higher-order mode OPO requires a complex setup, severely limiting the attainable squeezing and entanglement levels.

Direct mode conversion with fundamental mode squeezed light can avoid higher-order mode OPO

and nonlinear transformation. Nicolas Treps et al. transformed fundamental mode squeezed light into HG10, HG20, and HG30 modes squeezed light via deformable mirrors (DM) [27]. The limited number of pixels of a DM makes it unsuitable for generating complex spatial modes. Compared with the DM, the spatial light modulator (SLM) has more pixels and can finely control the light field. The amplitude and phase of the incident light can be simultaneously modulated by single or cascade SLMs [28-37]. Therefore, the SLM has received increasing attention as a mode conversion device. Marion Semmler et al. used a SLM to generate single-mode squeezing in LG and Bessel–Gauss (BG) modes of different orders, as well as an arbitrary intensity pattern [38]. The LG and BG modes were generated by shaping the spatial phase distribution of the light beam directly without touching the amplitude distribution. The efficiency was 0.15 for an arbitrary intensity pattern. However, higher-order HG modes and arbitrary complex amplitude distributed modes cannot be produced simultaneously with high efficiency and quality by single modulation.

In this paper, we demonstrate higher-order HG modes and arbitrary complex amplitude distributed squeezed state can high efficiency generated using a beam shaping system (BSS) on the fundamental mode squeezed state. The maximum mode conversion efficiency is 0.77. As a quantitative benchmark for the generated mode quality, we also analyze the mode purity by comparing the generated mode and corresponding theoretical standard mode.

## 2. Experimental setup

The experimental setup (Fig. 1) entails optical parametric amplification (OPA) squeezing, a BSS, and a purity measurement of higher-order modes squeezed light. First, the OPA is used to generate HG00 mode squeezed light. Second, the HG00 mode squeezed light is converted to higher-order modes squeezed light via cascade SLMs. Finally, the generated higher-order modes squeezed light are analyzed with respect to the quality of the spatial modes and the quantum noise reduction.

*2.1 OPA squeezing*

A continuous wave all solid state laser source emits both infrared at 1080 nm and green light at 540 nm. Part of the 1080 nm light is injected into the OPA as a seed beam. The seed beam is used to lock the OPA cavity [39]. Our OPA cavity is formed by a KTP crystal and a plano-concave mirror. The radius of curvature of the mirror is 20 mm, has a reflectance of 95% at 1080 nm and high transmittance at 540 nm. The seed beam is injected at the crystal surface, which is highly reflective (R>99.95%) at both 1080 nm and 540 nm. The OPA has a finesse of 120 with a free spectral range of 7.5 GHz. The OPA is pumped by 540 nm light, and the relative phase of the seed beam and pump beam is locked in the state of de-amplification.

The generated squeezed light was measured via ordinary homodyne detection [40]. The reflectance of the squeezed light on the beam splitter is much higher than the transmittance. In our setup, 98% of the squeezed light is reflected and 2% of the strong local beam is transmitted (Fig. 1(a)). By varying the phase of the local beam, different quadratures of the squeezed light can be measured. We assumed a perfectly coherent local beam and neglected the excess noise. We used a mode cleaner in the local beam path to ensure a coherent local beam within the detection band. The shot noise limit (SNL) was measured with the squeezed light input blocked. The squeezing output from the OPA was measured by scanning the phase of the local beam.

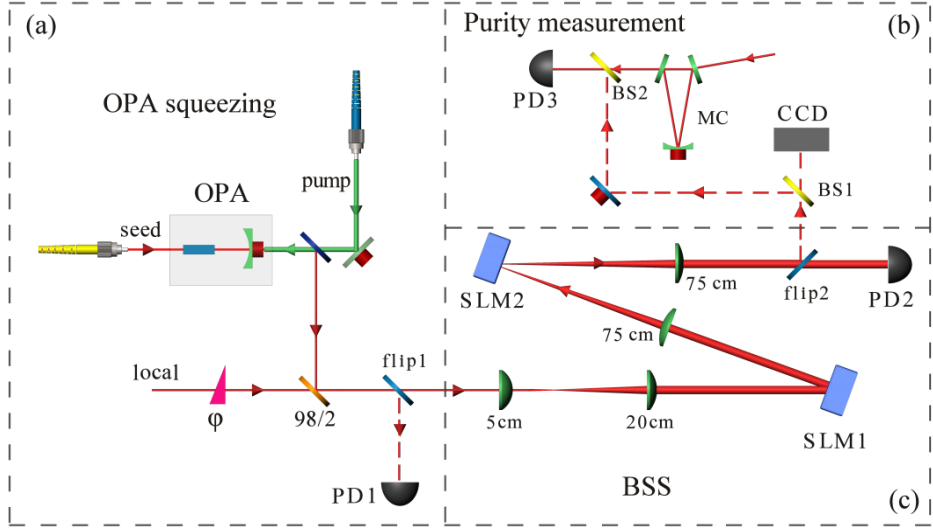

Fig. 1. Schematic of the experimental setup. The squeezed state in the HG00 mode of the OPA is first measured at PD1 with flip1. The BSS changes the spatial profile of the light, and the squeezing level is then measured at PD2. Flip2 is used to direct the generated modes for purity measurement. PD: photoelectric detector; HWP: half wave plate; BS: beam splitter; MC: mode cleaner.

*2.2 Higher-order mode generation*

The squeezed state is highly sensitive to optical loss. When a squeezed state experiences optical loss, it remains squeezed but the degree is reduced, limiting applications. A theoretically lossless method has realized both amplitude and phase modulation of the input beam with cascaded phase-only SLMs [32]. In contrast with existing techniques, the method theoretically allows an efficiency of almost 100%. It is possible to generate any desired light field distribution.

As shown in Fig. 1(c), the HG00 mode squeezed light is expanded by a telescope, which consists of two lenses with focal lengths of 5 cm and 20 cm separated by the sum of their focal lengths. We have taken the light output to be perfectly collimated with a waist of 5 mm. The collimated light is converted into higher-order spatial modes by the BSS. The BSS consists of two phase-only SLMs (Hamamatsu, X10648-03, pixel size = 20 um, 792×600 pixels) and two Fourier transforming lenses (focal length = 75 cm). Both elements are arranged in a 4f system. The light polarization (P polarization) corresponds to the working direction of the two SLMs.

The spatial amplitude and phase distributions can be programmed independently. This is achieved by diffracting the light from two phase-only SLMs located in conjugate Fourier planes. The amplitude distribution on SLM2 is created by SLM1, which can be iteratively optimized using the Gerchberg–Saxton (GS) algorithm [41]. However, a consequence of this optimization is that the field in the plane of SLM2 has a random phase. We correct the phase distribution through SLM2 by loading a phase correction hologram onto it. Altogether, this procedure generates the higher-order spatial modes in the target plane. Theoretically, we can obtain an arbitrary complex amplitude field.

For comparison, Fig. 2 shows two typical holograms loaded onto SLM1 and SLM2 to generate HG10 and HG50 modes. The hologram on SLM1 was obtained via 100 iterations of the GS algorithm. It is clear that the higher the mode order, the more complex the hologram is. Different gray values in the holograms represent different phases.

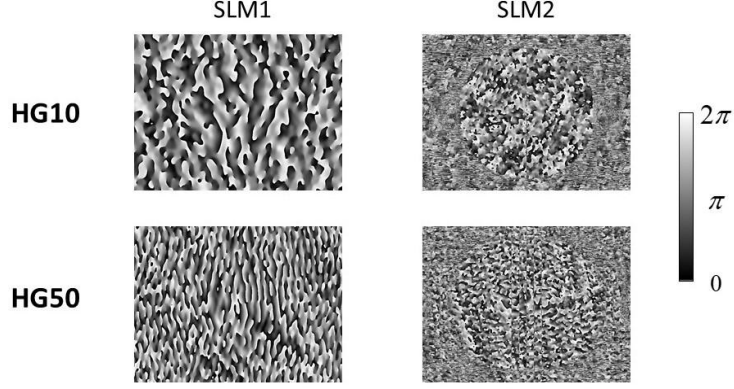

Fig. 2. Holograms loaded onto SLM1 and SLM2 for generating HG10 and HG50 modes.

*2.3 Higher-order-mode squeezed-light measurement*

The generated higher-order mode squeezed light was analyzed with respect to the quantum noise reduction and the quality of the spatial modes. PD2 was placed in the target plane, and squeezing was measured by scanning the phase of the local beam. For each generated mode, we used two separate homodyne detectors (PD1 and PD2 in Fig. 1(a) and (c)) to measure the squeezing before and after the BSS.

Because of the limited SLM resolution, it is impossible to generate the standard intensity and phase distribution, and thus perfect target mode, in practice. The purity of higher-order modes is defined by the visibility of interference between the generated mode and the standard mode of the same order. A high-finesse mode cleaner is used as a standard Gaussian mode selector. This cavity is seeded with a general higher-order mode and locks it in resonance with selected higher-order HG modes. When flip2 is present, we measure the degree of interference with standard modes. A charge-coupled device (CCD) (Hamamatsu, C10633) is used to capture the intensity distribution of the generated modes (Fig. 1(b)).

## 3. Results and discussion

*3.1 Higher-order modes squeezing*

In our setup, the OPA threshold is 400 mW, with a pump power of 280 mW and an injected seed beam of 5 mW. Before the BSS, we typically observed $-5.22\pm 0.20$ dB squeezing of the HG00 mode at PD1 with flip1, as shown in Fig. 3. Trace 1 corresponds to the SNL, which is obtained by blocking the signal beam. Trace 2 corresponds to the quantum noise levels of the HG00 mode squeezed state with the local beam phase scanned. All data are normalized using the SNL level.

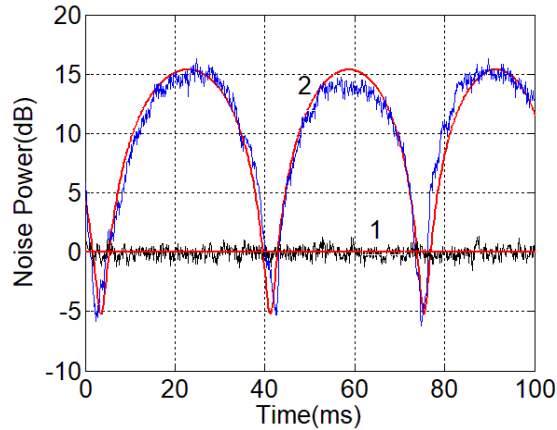

Fig. 3. OPA squeezing. The measurement parameters of the spectrum analyzer are: RBW: 100 kHz; VBW:

100 Hz; analysis frequency: 3 MHz.

Figure 4 shows all the squeezing for the generated modes. Trace 1 corresponds to the SNL, which is obtained by blocking the squeezed light. Trace 2 corresponds to the quantum noise levels of the generated mode squeezed state with the local beam phase scanned. Here, a noise reduction of $-2.65\pm0.19$ dB below the SNL for the HG50 mode can be seen. To our knowledge, this is the first measurement of squeezing at such a high order HG mode. For an optical pattern of our laboratory initials QMC, the noise reduction of $-2.36\pm0.21$ dB below the SNL.

The difference in squeezing levels between PD1 and PD2 can be accounted for by losses in power for each of the different transformations. The approximately 20% loss comes from the absorption and imperfect diffraction efficiency of the SLMs. The additional 2% loss is consistent with the number of optical elements in the beam path and the specifications of their coatings. The total efficiency is reduced further by a few percent for modes with higher orders owing to the limited aperture of the optical element, resulting in high spatial frequency losses, and the more complex the distributed mode, the more obvious this phenomenon is. The losses for the higher-order mode squeezing can be expressed as

$$V_{out} = \eta \cdot V_{in} + (1-\eta) \cdot V_{vac} , \qquad (1)$$

where $\eta$ is the mode conversion efficiency, $V_{in}$ and $V_{out}$ represent the variances of the input and output beams of the squeezed quadrature, and $V_{vac}$ is the vacuum variance. In our setup, the maximum mode conversion efficiency is 0.77, which is calculated from the HG10 mode squeezing. For QMC, the mode conversion efficiency is 0.6. This shows that the BSS gives access to high squeezing levels and high efficiency in arbitrary complex distributed modes.

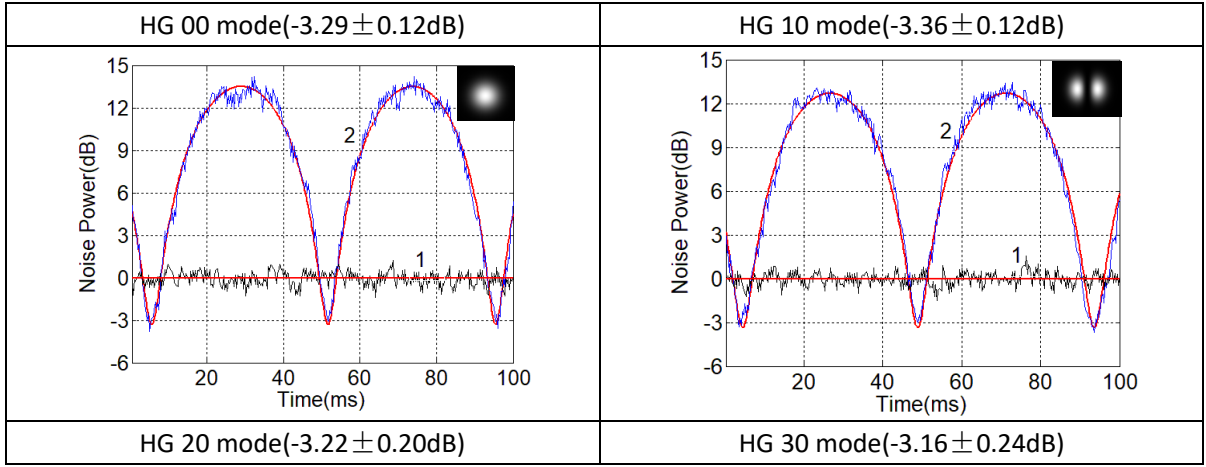

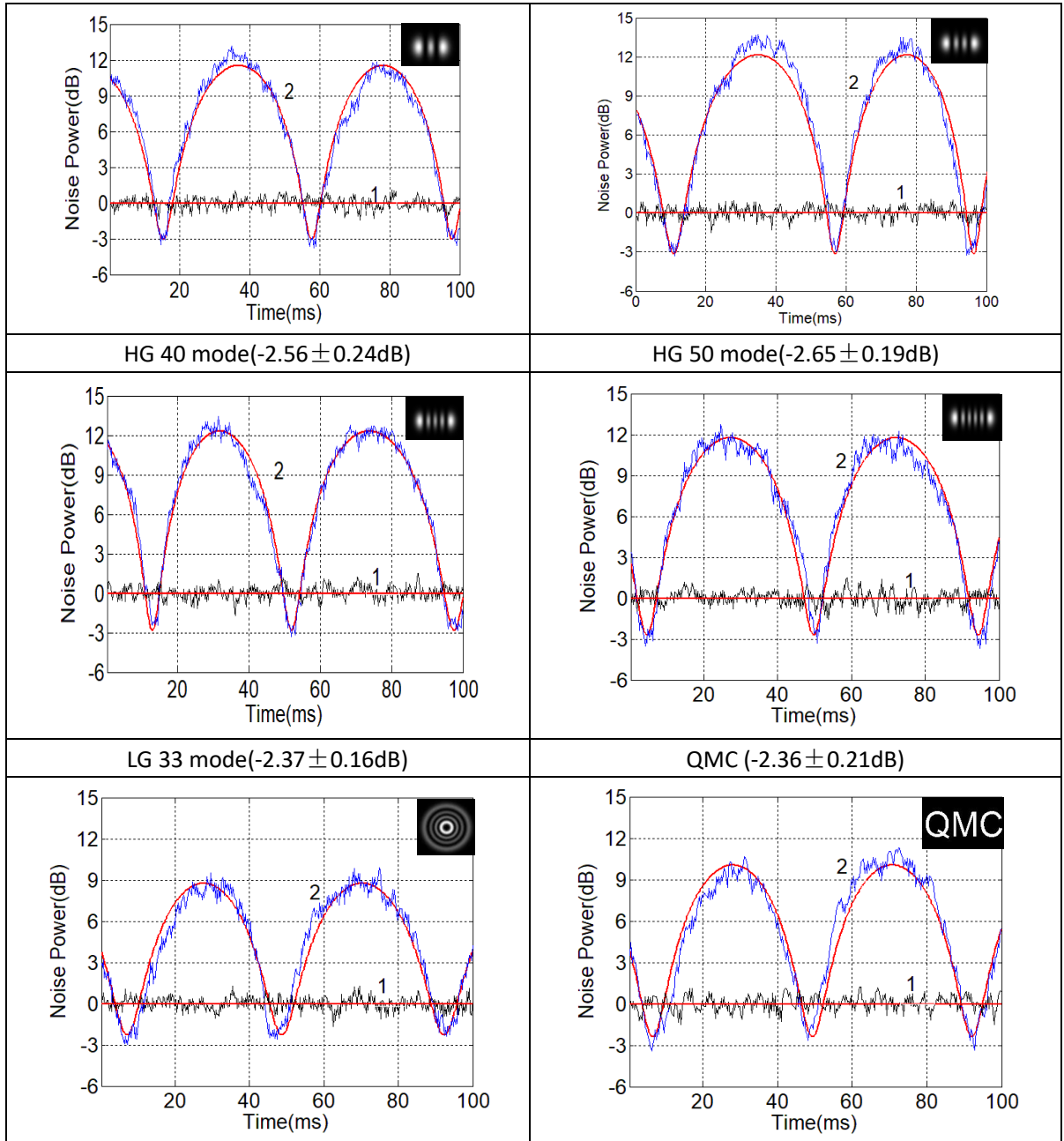

Fig. 4. Squeezed spatial modes. The measurement parameters of the spectrum analyzer are: RBW: 100 kHz; VBW: 100 Hz; analysis frequency: 3 MHz.

*3.2 Higher-order modes purity analysis*

The first five order HG modes were generated by the BSS, and then the LG33 mode and QMC are taken as an example in order to show the manipulation of arbitrary complex amplitude field. For QMC, the phase and intensity is uniform. We measured the interferograms, i.e., the interference between the generated modes and a Gaussian beam reference with a waist of 6 mm, to identify the phase distribution of the generated modes. The experimental results are shown in Fig. 5. The intensity distribution and interferograms agree well with the theoretical ones.

| Theoretical intensity | Theoretical interferogram | Experimental intensity | Experimental interferogram |
|---|---|---|---|
| | | | |
| | | | |
| | | | |
| | | | |
| | | | |
| | | | |
| | | | |
| | | | |

Fig. 5. Theoretical and experimental modes and interferograms.

The mode purity is calculated as an inner product between the generated mode and the theoretical standard mode, given by the equation [42]

$$P = \iint |b(x,y)c^*(x,y)|^2 dxdy \bigg/ \iint |b(x,y)|^2 |c(x,y)|^2 dxdy ,\qquad(2)$$

where $b(x,y)$ is the mode generated by the BSS, and $c(x,y)$ is the theoretical standard mode. We quantitatively analyzed the purity of the generated modes. First, the mode purity was obtained from the intensity distribution and interferograms via intensity analysis method [43]. The phase distribution can be obtained from the interferograms in Fig. 4 to calculate the mode purity via Eq. (2).

Next, the mode purity was measured using the visibility of interference with the standard mode of the same order (Fig. 1(b)). Because it is difficult to produce the standard LG33 mode and the QMC pattern in our experiment, we did not obtain the purity of the LG33 mode or QMC [27]. Figure 6 exhibits the behavior of output mode purity via calculated results and measured results. The calculated and measured results are similar, and the generated mode purity decreases as the mode order increases.

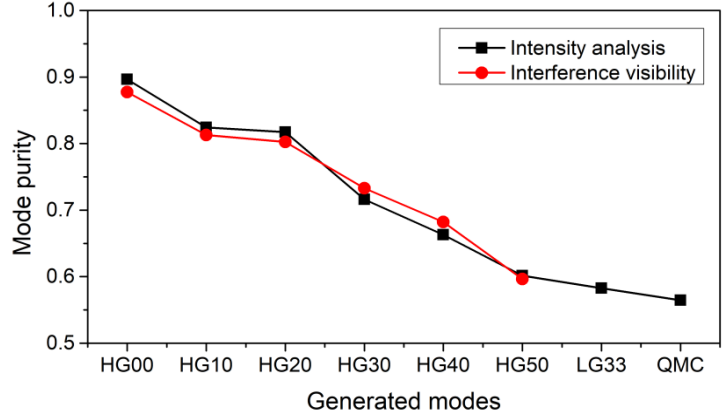

Fig. 6. Purity of the generated modes obtained from intensity analysis and interference visibility.

Remarkably, one may still observe some deviations in the experimental results compared to theories, especially for higher-order (more complex distributed) modes. This might be explained as follows. First, the SLM suffers from crosstalk between adjacent pixels, inducing errors between the calculated hologram and the one actually loaded on the SLM. This results in inevitable deterioration of the generated mode quality, especially for higher-order (more complex distributed) modes [44]. As seen in Fig. 2, the holograms for the HG50 mode are more complex than those for the HG10 mode. Next, the high spatial frequency is lost owing to the limited aperture of the optical element, causing deviation of the phase distribution. A crucial point for achieving good mode quality is accurate alignment of SLM2. A higher mode order leads to a high spatial frequency and thus a high sensitivity to the alignment of SLM2. The deviation in phase distribution more affects the phase correction for SLM2 [32]. Finally, the presence of a small unmodulated part of the light further diminishes the mode quality. However, this effect might be minimized by future technical improvements to the SLM that could increase diffraction efficiency [45]. We expect that a mode purity reaching 0.9 could be achieved by optimizing the above factors.

## 4. Conclusion

We have shown that the BSS can transfer squeezing from the fundamental mode to an arbitrary complex amplitude distributed mode with a high efficiency of 0.6. With this method, different spatial modes can be generated simply by applying different holograms on the SLMs. Our system does not disrupt the quantum properties of the light.

High efficiency mode conversion can be applied in multiplex quantum information processing with structured light, to solve the problem of low detection efficiency for quantum states in high-order spatial modes [46]. The generated high-order spatial mode squeezed state has promising application in quantum metrology such as super resolution quantum imaging [3], mitigating thermal noise [9] or mode matching loss [47] in LIGO interferometry, and realizing CV parallel quantum information protocols based on spatial multimode squeezed states [11].

**Funding**

Key Project of the Ministry of Science and Technology of China (2016Y-FA0301404); National Natural Science Foundation of China (NSFC) (91536222, 11674205); Program for OIT of Shanxi and Shanxi 1331 Project.

**Disclosures**